\documentclass[prb,twocolumn]{revtex4}
\usepackage{graphicx}
\def\eps{\epsilon_{\rm eff}}
\def\epsm{\epsilon_{\rm m}}
\def\cl{c_{\rm light}}
\def\be{\begin{equation}}
\def\ee{\end{equation}}

\begin{document}

\title{Absorption losses in periodic arrays of thin metallic wires}

\author{P. Marko\v{s}$^*$ and C. M. Soukoulis}
\affiliation{Research Center of Crete, 71110 Heraklion, Crete, Greece\\
and\\
Ames Laboratory and Department of Physics and Astronomy, Iowa State University, Ames, Iowa 50011}

\begin{abstract}
We analyze the transmission and reflection of the electromagnetic wave 
calculated from transfer matrix simulations
of periodic arrangements of thin metallic wires. The effective permittivity and the absorption is
determined. Their dependence on the wire thickness and the conductance of the metallic wires is studied.
The cutoff frequency or effective plasma frequency is obtained  and is compared with analytical
predictions. It is shown that the periodic arrangement of wires exhibits a frequency region 
in which the real part of the permittivity is negative while its imaginary part is very small.
This behavior is seen for  wires with thickness as small as 17 $\mu$m with a lattice constant of 3.33 mm.
\end{abstract}

\maketitle

Rapidly increasing interest in left-handed metamaterials (LHM) (for a recent review, see \cite{MSnato}) raised
some interesting questions about the electromagnetic (EM) properties of composites which contain
very thin metallic components. The most simple example of such a composite is a periodic array of 
thin metallic wires. Pendry {\it el al.} \cite{Pendry}  predicted that such system behaves
as a  high pass filter with an effective permittivity
\be\label{epseff}
\eps=1-\frac{f^2_p}{f^2+2i\gamma f}.
\ee
In Eq.  \ref{epseff}, $f_p$ is the effective  plasma frequency, or
cutoff frequency \cite{Sigalas}, and $\gamma$ is the damping factor.
Various theoretical formulas were derived for the 
dependence of the plasma frequency  on the lattice period $a$ and the wire radius 
$r$.
Pendry {\it el al.} \cite{Pendry} obtained that
\be\label{f-pendry}
f_p^2=\frac{\cl^2}{2\pi a^2\ln(a/r)},
\ee
Shalaev and Sarychev \cite{Shalaev} obtained that
\be\label{f-shalaev}
f_p^2=\frac{\cl^2}{2\pi a^2[\ln(a/\sqrt{2}r)+\pi/2-3]},
\ee
while 
Maslovski {\it et al.} \cite{Maslovski} that
\be\label{f-maslovski}
f_p^2=\frac{\cl^2}{2\pi a^2[\ln a^2/4r(a-r)]}.
\ee
In Eqs. \ref{f-pendry}-\ref{f-maslovski}, $\cl$ is the velocity of light in vacuum. 

Periodic arrangements  of thin metallic wires are  used as a  negative-$\epsilon$ medium
\cite{Brown,Rotman}
 in the
left-handed structures \cite{Smith,Shelby,Claudio}. It is therefore important to understand 
how the  electromagnetic response  - 
not only the effective plasma frequency, but also the  factor $\gamma$ - 
depends on the structural parameters of the wire  system. 
Recently, 
Ponizhovskaya {\it el al.} \cite{PGN} claimed that for small wire radius, the absorption in the wire system
is so large that the transmission losses  do not allow any propagation of EM 
wave in left-handed structure.
Very low transmission, measured in the original  experiments on the LHM, \cite{Smith,Shelby}  seemed to agree with
their pessimistic conclusion. 
However, it is not clear why the transmission was so low in the original experiments.
Recent  experimental measurements  \cite{Claudio,Ekmel} established that the transmission
of LHM could be as good as in the right-handed systems.

Our aim in this Letter is to study numerically   how the effective permittivity of the 
periodic arrangement of metallic wires depends
on the wire radius and on the conductance of the wires. We
present  results for the real ($\eps'$) and imaginary ($\eps''$) part of the
effective permittivity of the wire medium, estimate the transmission losses
and the plasma frequency and compare our results with the analytical formulas given in Eqns.
\ref{epseff}-\ref{f-maslovski}.

In our numerical simulations we use the transfer matrix method (TMM).
Details of the method are given elsewhere \cite{MS-1}. Here we only  point out the main advantage of
the TMM, namely that it  gives  directly the transmission $t$, reflection $r$ and absorption 
$A=1-|t|^2-|r|^2$ of the EM plane wave passing through the system. Contrary to this, 
in the FDTD method, which was used in Ref. \cite{PGN},
one obtains $t$ and $r$ form the time development of the wave packet, which is much
more complicated and probably  also less accurate.
To be able to obtain $\eps$, one also needs  the phase of $r$ and $t$, in addition to their amplitudes.
This is also easily achieved in the TMM.

From the obtained data for the transmission and reflection, we calculate the effective permittivity of the system.
\cite{MS-2}  The refraction index is given by the  equation
\be
\cos(nkL)=\frac{1}{2t}\left[1-r^2+t^2\right].
\ee
Since  we do not expect any magnetic response, we fixed the value of the permeability  to be $\mu\equiv 1$.
The permittivity is then found as 
$\eps=n^2$.

The way we discrete the space in the TMM  might control the accuracy of our results. 
To test how discretization   influences our results, 
we repeated the numerical simulation  for different  discretizations. 
The wire is represented as a rectangular  with square cross section $2r\times 2r$, $r$ being the ``wire radius''.
The obtained results for the 
effective permittivity $\eps$ (both real and imaginary part)
are almost  independent of the discretization procedure. 

Figure 1 shows how the effective permittivity depends on the 
wire radius. We analyzed four  different wire arrays with period $a=3.33$~mm.
For all of them  the real part of the  effective permittivity is negative and could be fitted by Eq. (\ref{epseff}).
This enables us to obtain easily the plasma frequency. 
Only for the smallest wire thickness studied  ($17\times 17\mu$m) one  gets a relatively large $\eps''$
for small frequencies. In this region relation (\ref{epseff}) is not valid.  Nevertheless, for frequencies larger than
5 GHz, $\eps''$ is small and $\eps'$ negative.

The right lower panel of fig. 1 shows data for  wires with  the cross section of $17\times 300\mu$m.
These parameters were used  in the experiment of Shelby {\it et al.} \cite{Shelby}. We again see that 
the formula given by Eq. \ref{epseff}  qualitatively agree with our data. Thus, there is no doubt that
this array of wires really produces a medium with  negative  $\eps'$, which then
can be used in the creation of  the left handed systems.

Figure 2 compares our data for plasma frequency with the analytical 
formulas given by Eqns. \ref{f-pendry}-\ref{f-maslovski}.
Accepting some uncertainty in the estimation of the plasma  frequency from the numerical data, we can
conclude that for thin wires our data agree  with the theoretical formula (\ref{f-shalaev}) 
of Shalaev and Sarychev. \cite{Shalaev}
For thicker wires, our results are in agreement with 
the  formula (\ref{f-maslovski}) of Maslovski {\it et al.} \cite{Maslovski}

We also study the role of the conductance of the metallic wires. 
In the simulations shown in
fig. 1, we consider the metallic permittivity to be $\epsm=(-3+588~i)\times 10^3$. We are aware that this 
value of $\epsm$ is 
smaller than the permittivity of realistic metallic wires: for instance for copper 
 $\epsm\approx 5\times 10^7~i$, as follows from the relation between the permittivity and the conductance
\cite{Jackson} (the conductivity of copper is  $\sigma\approx 5.9\times 10^7(\Omega{\rm m})^{-1}$).
Our data in fig. 1 therefore underestimate losses, because 
transmission losses are smaller for higher values of the metallic permittivity \cite{Pendry-2}.
This is clearly shown in fig. 3 where we  present 
the ratio $\kappa=|\eps''/\eps'|$ {\it vs} frequency for 
two systems which differ only in the value of the imaginary part of the metallic permittivity.
Fig. 3 shows also the frequency dependence of the 
absorption as obtained from the numerical simulations. The absorption also exhibits a maximum
in the neighbor of the plasma frequency. 

Fig. 1 also confirms that $\eps''$ increases when the wire radius decreases. For instance, the
parameter $\gamma$ is only 0.003 GHz for $r=100\mu$m. As we will see below, (fig. 4) 
$\gamma$ increases up to 1.2 GHz when the wire radius decreases to 15$\mu$m. Nevertheless, even for
very thin wires,
losses are much less than what was claimed in Ref. \cite{PGN}.  As it is shown in fig 1,
also an array of  wires with  thickness $17\times 17\mu$m creates a negative-$\epsilon$ medium.
As this is in 
strong contrast with the  results of Ref. \cite{PGN}, we decided to study  exactly the same system as
that of \cite{PGN}. Results of our simulations are shown in fig. 4. Although
such systems are not used in experimental arrangements of the LHM, our results give a comparison between
two different numerical treatments. Our data again clearly show that 
$\eps'$ is {\it negative} for $f<f_p$.  As it is shown in the inset, the transmission losses 
are also small. 

We believe that the present data are more accurate, than those published previously in \cite{PGN},
not only because they agree with the theoretical analysis but also because the TMM gives 
the reflection and its phase straightforward. This is important because the main difference
between our results and those of \cite{PGN} seems to be in the estimation of reflection $R=|r|^2$.
When comparing our data for absorption, given in the inset of fig.  
4 with those given in fig 2b  of Ref. \cite{PGN}, we   see that  our absorption is much less
than that estimated in  Ref. \cite{PGN}. 

\medskip

In conclusion, we analyzed numerically the transmission properties of a periodic
arrangements  of thin metallic wires.
From the transmission and reflection data we  calculate the effective permittivity and plasma frequency, which agrees 
qualitatively with theoretical predictions. Both the effective permittivity
and the absorption data confirm that the array of thin metallic wires, used in the recent experiments
on the left-handed metamaterials,
indeed behaves as  a negative permittivity medium with low losses.

\medskip
\noindent{\bf Acknowledgments:} 
This work was supported
by Ames Laboratory (Contract. n. W-7405-Eng-82). Financial support of DARPA,
NATO (Grant No. PST.CLG.978088), APVT (Project No. APVT-51-021602) 
and EU project DALHM are also acknowledged.

\begin{figure}
\centerline{\scalebox{0.45}{\includegraphics{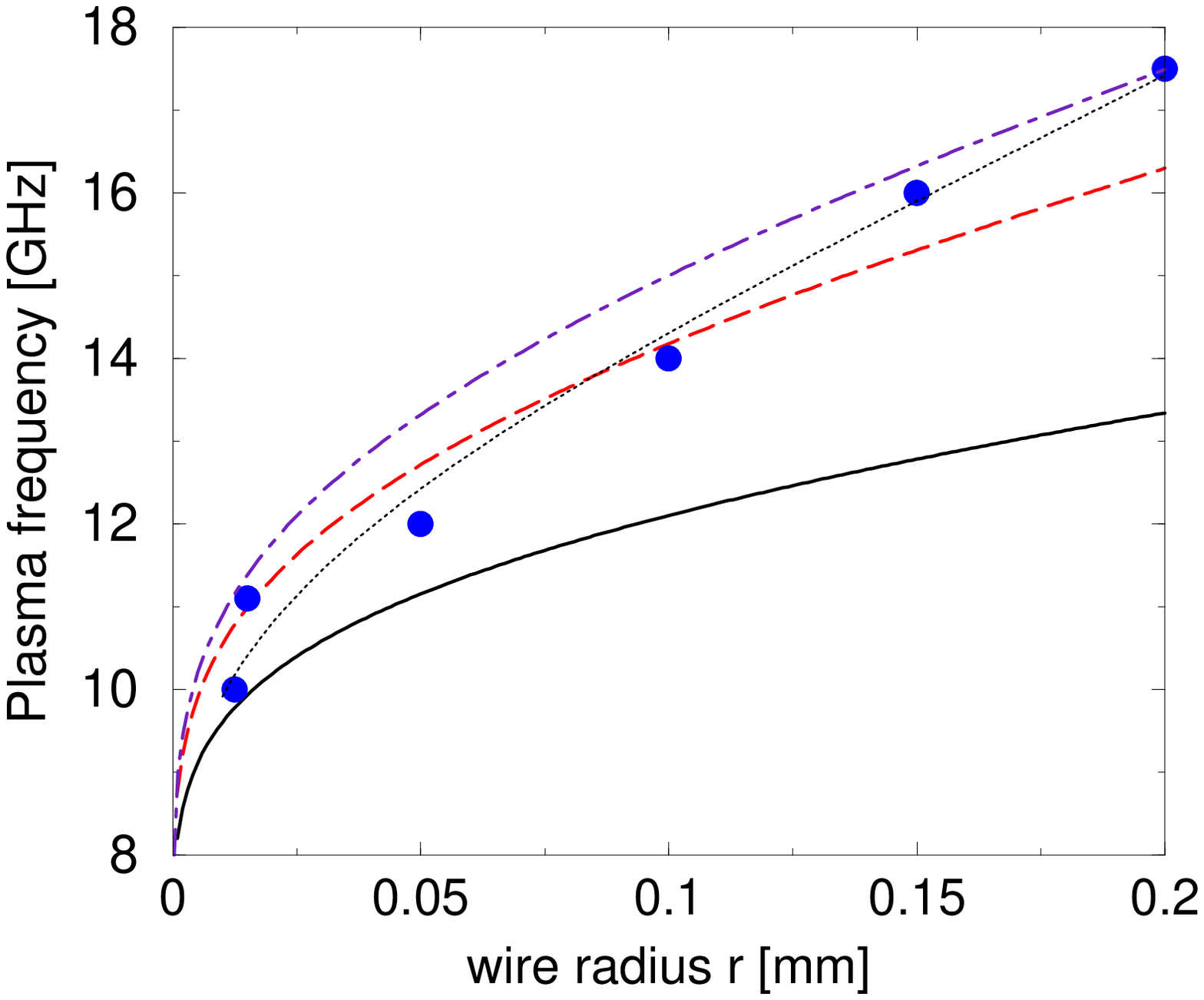}}}
\caption{Effective permittivity as a function of frequency for various shapes of the metallic wires.
The lattice period in all cases is $a= 3.33$ mm. We used  the metallic permittivity $\epsm=(-3+588~i)\times 10^3$.}
\end{figure}

\begin{figure}
\centerline{\scalebox{0.45}{\includegraphics{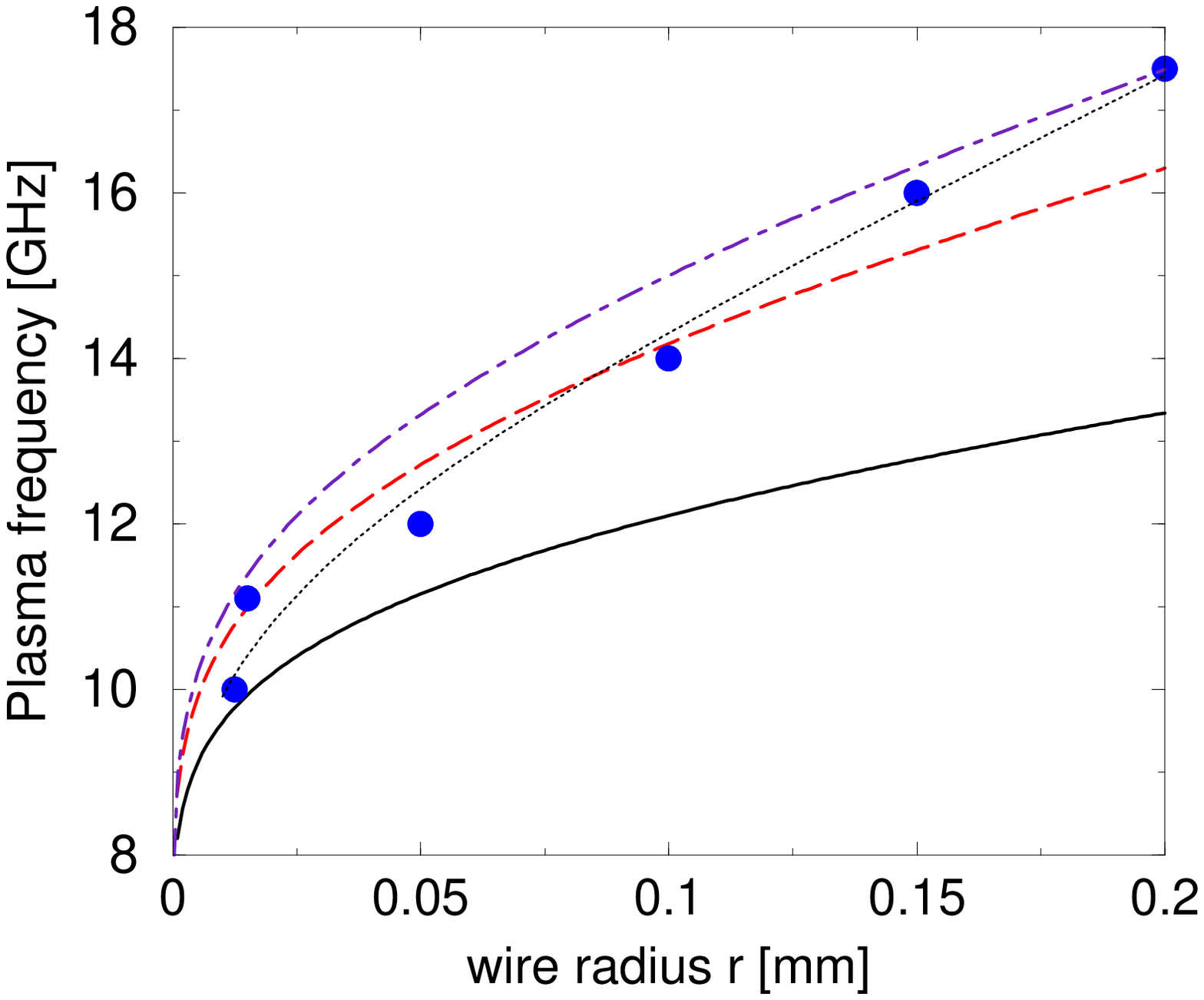}}}
\caption{Plasma frequency as a function of the wire radius. The lattice constant is
$a=5$ mm. Solid, dashed and dot-dashed  line is the result
of Pendry {\it et al.} (\ref{f-pendry}), Shalaev and Sarychev (\ref{f-shalaev}) 
and Maslovski {\it et al.} (\ref{f-maslovski}), respectively. 
Dot line is a fit of our data to 
the function $f_p=a_0/\sqrt{\ln(a_1/r)}$ with parameters $a_0=20.9$ and $a_1=0.84$. 
}
\end{figure}

\begin{figure}
\centerline{\scalebox{0.45}{\includegraphics{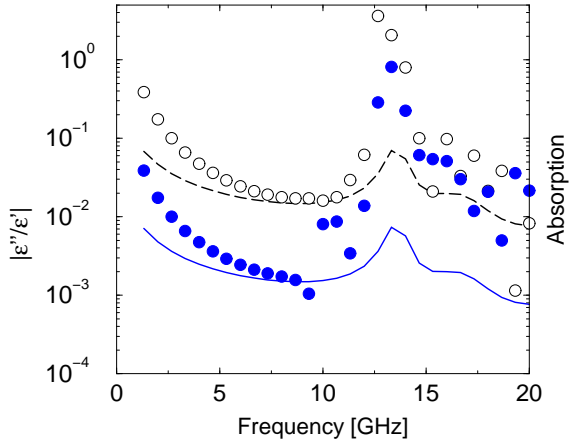}}}
\caption{Ratio $\kappa=|\eps''/\eps'|$ for a lattice of wires with radius $50\mu$m.
Metallic permittivity is $\epsm=(-3+588~i)\times 10^3$ (open circles) and
$\epsm=(-3+5~880~i)\times 10^3$ (full circles). Dashed (solid) line is absorption for
the corresponding systems obtained numerically by the TMM.
These numerical results confirm that losses are smaller for higher metallic permittivity
and that the value of the plasma frequency, estimated approximately from the position of the
maximum of $\kappa$, does not depend on the value of the metallic permittivity.}
\end{figure}

\begin{figure}
\centerline{\scalebox{0.45}{\includegraphics{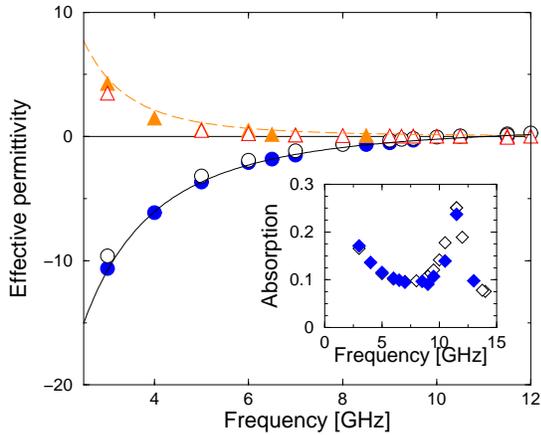}}}
\caption{Effective permittivity for a lattice of thin metallic wires. 
The wire radius is $r= 15\mu$m and the  lattice constant is $a=5$ mm. The metallic permittivity
is $\epsm=-2000+10^6~i$. Two different 
discretizations are used with mesh sizes of $30\mu$m (open symbols) and $15\mu$m (full symbols).
The solid and dashed lines are fit to Eq. \ref{epseff} with $f_p=11.1$ GHz and 
$\gamma=1.2$ GHz. The length of the system
was up to 60 unit lengths for open symbols and 10 unit lengths for full symbols.
Inset shows the numerically calculated  absorption as a function of frequency.}
\end{figure}

\end{document}